\documentclass{article}

\usepackage{arxiv}

\usepackage{cite}
\usepackage{amsmath,amssymb,amsfonts}
\usepackage{algorithm}
\usepackage{algpseudocode}
\usepackage{graphicx}
\usepackage{textcomp}
\usepackage{subfigure}
\usepackage{wrapfig}
\usepackage{ulem}
\usepackage{authblk}

\usepackage[utf8]{inputenc} 
\usepackage[T1]{fontenc}    
\usepackage{hyperref}       
\usepackage{url}            
\usepackage{booktabs}       
\usepackage{amsfonts}       
\usepackage{nicefrac}       
\usepackage{microtype}      
\usepackage{doi}

\title{Design and Implementation of the Illinois Express Quantum Metropolitan Area Network}

\author[1,2,*]{Joaquin Chung}
\author[4]{Ely M. Eastman}
\author[3,4]{Gregory S. Kanter}
\author[7] {Keshav Kapoor}
\author[5,6]{Nikolai Lauk}
\author[7]{Cristián Peña}
\author[7]{Robert Plunkett}
\author[5,6,8]{Neil Sinclair}
\author[4]{Jordan M. Thomas}
\author[5,6]{Raju Valivarthi}
\author[5,6]{Si Xie}
\author[1,2,9]{Rajkumar Kettimuthu}
\author[4,10]{Prem Kumar}
\author[7]{Panagiotis Spentzouris}
\author[5,6]{Maria Spiropulu}

\affil[1]{Data Science and Learning Division, Argonne National Laboratory, 9700 S Cass Ave. Lemont, IL. 60439, USA}
\affil[2]{The University of Chicago Consortium for Advanced Science and Engineering, 924 E 57th St, Chicago, IL 60637, USA}
\affil[3]{NuCrypt LLC, 1840 Oak Ave, Evanston, IL 60201, USA}
\affil[4]{Center for Photonic Communication and Computing, Department of Electrical and Computer Engineering, McCormick School of Engineering and Applied Science, Northwestern University, 2145 Sheridan Road, Evanston, IL 60208-3118, USA}
\affil[5]{Division of Physics, Mathematics and Astronomy, California Institute of Technology, 1200 E. California Blvd., Pasadena, CA 91125, USA }
\affil[6]{Alliance for Quantum Technologies (AQT), California Institute of Technology, 1200 E. California Blvd., Pasadena, CA 91125, USA}
\affil[7]{Fermi National Accelerator Laboratory, Kirk Rd and Pine St, Batavia, IL 60510, USA}
\affil[8]{John A. Paulson School of Engineering and Applied Sciences, Harvard University, 29 Oxford St., Cambridge, MA 02138, USA}
\affil[9]{The Northwestern Argonne Institute of Science and Engineering (NAISE), 2205 Tech Drive Suite 1-160, Evanston, IL 60208, USA}
\affil[10]{Department of Physics and Astronomy, Northwestern University, 2145 Sheridan Road, Evanston, IL 60208-3112, USA}

\affil[*]{Corresponding author: chungmiranda@anl.gov}

\begin{document}
\maketitle

\begin{abstract}
The Illinois Express Quantum Network (IEQNET) is a program to realize metropolitan scale quantum networking over deployed optical fiber using currently available technology. 
IEQNET consists of multiple sites that are geographically dispersed in the Chicago metropolitan area. 
Each site has one or more quantum nodes (Q-Nodes) representing the communication parties in a quantum network. 
Q-Nodes generate or measure quantum signals such as entangled photons and communicate the measurement results via standard, classical signals and conventional networking processes.
The entangled photons in IEQNET nodes are generated at multiple wavelengths, and are selectively distributed to the desired users via transparent optical switches. 
Here we describe the network architecture of IEQNET, including the Internet-inspired layered hierarchy that leverages software-defined networking (SDN) technology to perform traditional wavelength routing and assignment between the Q-Nodes. 
Specifically, SDN decouples the control and data planes, with the control plane being entirely implemented in the classical domain. We also discuss the IEQNET processes that address issues associated with synchronization, calibration, network monitoring, and scheduling. 
An important goal of IEQNET is to demonstrate the extent to which the control plane classical signals can co-propagate with the data plane quantum signals in the same fiber lines (quantum-classical signal ``coexistence''). 
This goal is furthered by the use of tunable narrow-band optical filtering at the receivers and, at least in some cases, a wide wavelength separation between the quantum and classical channels. 
We envision IEQNET to aid in developing robust and practical quantum networks by demonstrating metro-scale quantum communication tasks such as entanglement distribution and quantum-state teleportation.   
\end{abstract}

\keywords{Quantum networks, metropolitan area, Q-LAN, Q-MAN}

\section{Introduction} \label{sec:intro}

The full exploitation of quantum computing and quantum sensing is expected to require fully networked and distributed solutions, as is robustly shown in the classical case. The advantages presented by deploying such quantum systems come along with special challenges to the transmission of information that must be addressed to realize the full potential of networking.  For example, quantum signals are very sensitive to loss and even minuscule levels of added noise.  The no-cloning theorem prevents us from using approaches commonly employed in classical networks to mitigate such problems since it forbids the creation of identical copies of an arbitrary unknown quantum state.  As a result, more sophisticated approaches must be employed to protect the quantum information we want to transfer (for example, through indirect channels utilizing distribution of quantum entanglement and teleportation).

While quantum optical networks will eventually be used to interconnect quantum computers, for example via quantum transduction~\cite{lauk2020perspectives, lambert2020coherent}, it is worth to mention that they provide some important near term advantages such as enhancing the security of communication via quantum cryptography and improving the sensitivity of measurements via distributed quantum sensing.
While it is desirable to leverage as much of the already deployed  fiber-optic infrastructure as possible, because of the quantum information transmission specific challenges we discussed above, we expect quantum network design to require major paradigm shifts from classical networks, as well as custom engineering for its systems.  At the same time, it is imperative that quantum networks are able to co-exist and, eventually, interoperate with classical networks, so the design must incorporate as much compatibility with conventional networks as is practical.



The Illinois Express Quantum Network (IEQNET) is a program for developing metro-scale, repeaterless quantum networking over deployed optical fiber infrastructure. While new technologies, such as quantum repeaters, will be needed to realize the long-term goals of quantum networking, IEQNET focuses on leveraging currently available technology (with provision for future upgrades as technology advances), 
to develop architecture and systems, and test and improve them through a co-design process with deployment of metropolitan area quantum networks. 

Two of the prime needs in quantum information systems are distributing entanglement and subsequently using the entanglement to perform tasks like quantum key distribution (QKD) and quantum teleportation. These needs should be met while simultaneously allowing higher-power classical signals to share the same fiber, both for the purposes of enabling communications to support quantum applications and for independent coexistence of high data-rate classical channels. In this paper we introduce our architecture to realize these functions over a metro-scale network, focusing on simplicity for entanglement distribution, and discuss some of the major control and management issues that need to be addressed to enable reliable network operation. While there have been several proof-of-concept demonstrations of deployed quantum communications and networking, over both free space and fiber, in various locations around the globe (see references~\cite{boaron2018secure,elliott2007darpa,liao2017satellite,peev2009secoqc,pirandola2020advances,sasaki2011field,ursin2007entanglement,Joshi2020} for an overview), there is a need to advance entanglement-based technologies to be more integrated into the (classical) networking framework beyond that of previous demonstrations. This includes scaling to more users, reaching longer link distances, allowing coexisting quantum and classical data channels on the same fiber links, testing quantum protocols in a realistic network setting, e.g., using transparent optical switches and other optical networking components, as well as optimizing and automating the software control of network operations such as synchronization between nodes.
\section{Related Work} \label{sec:related-work}
With a long-term vision of designing a quantum internet, researchers are proposing network architectures, roadmaps, and protocol stacks.
In 2016, Muralidharan et al. in~\cite{muralidharan2016optimal} proposed three generations of quantum repeater networks for optimal quantum communications that go from entanglement distribution and swapping without quantum error correction in the first generation to fully error-corrected logical qubits being transmitted in a hop-by-hop basis in the third generation.
Later in 2018, Wehner, Elkouss, and Hanson~\cite{wehner2018quantum} proposed a six stages roadmap for the development of a quantum internet that starts with trusted repeater networks for QKD applications and ends with a quantum internet capable of interconnecting quantum computers that run complex distributed algorithms.
In the context of such roadmap, IEQNET can be classified as a Stage 3 network or entanglement distribution network (without quantum memories).
Long et al.~\cite{long2022qsdc} further extended this roadmap with an intermediate stage to provision for quantum secure direct communication (QSDC) in the absence of quantum memories.
More recently in 2022, Van Meter et al.~\cite{van-meter2022qi} proposed a quantum internet architecture based on the Quantum Recursive Network Architecture (QRNA) and RuleSet-based connections establishment via two-pass connection setup.
Similarly, several quantum internet protocol stack proposals have emerged in recent years.
We refer the readers to~\cite{ILLIANO2022109092} for a comprehensive survey.

Taking a bottom-up approach, other research groups have recently demonstrated early prototypes of campus- and metropolitan-scale quantum networks.
For instance, Wengerowsky et al.~\cite{wengerowsky2018entanglement} demonstrated an entanglement-based, wavelength-multiplexed quantum communication network for QKD.
With the same aim of supporting QKD applications, Martin et al.~\cite{martin2019sdn-qkd} deployed the Madrid Quantum Network using a software-defined networking (SDN) control plane that integrates classical and quantum channels.
While these demonstrations may show architectural components similar to IEQNET, our design considers applications beyond QKD such as quantum teleportation and requires coexistence with classical network traffic.
Taking a more general approach, Pompili et al.~\cite{Pompili2022} experimentally demonstrated entanglement delivery using the quantum network protocol stack proposed in~\cite{dahlberg2019}.
Similarly, Alshowkan et al.~\cite{PRXQuantum.2.040304} demonstrated remote state preparation (RSP) on a reconfigurable quantum local area network that uses flexible-grid bandwidth allocation over deployed fiber.
Tessinari et al.~\cite{Tessinari21} experimentally proved the coexistence of a bright 100~Gbps classical communication signal with multiple single-photon level entanglement channels over a fiber optical quantum network using wavelength multiplexing.
\section{Design Considerations} \label{sec:design}

Despite the fact that quantum networks are not yet fully developed, 
it is expected that most quantum applications will require networks that support entanglement distribution to enable teleportation for quantum communications, in order to protect the quantum information. Since quantum network components essential for network deployment beyond the metropolitan scale (e.g., quantum repeaters) are currently unavailable, it is essential to advance quantum network architecture and develop quantum network technologies independently of the availability of such components.  One such approach is to develop entanglement-based quantum network architecture and implementations that employ software-defined networking (SDN) technology~\cite{adv-sdn-survey}, together with transparent optical switches to implement network functions.  Such an architecture design allows the network to establish lightpaths among quantum nodes or between quantum nodes and entangled photon sources via traditional (classical) wavelength routing and assignment approaches used in transparent optical networks. In transparent optical networking, a lightpath is a path between two nodes in the network in which light passes through unmodified.
This quantum network architecture approach is inspired by the vision of universal transparent quantum networks detailed in the quantum networks for science workshop report~\cite{thomasreport}.  It is also desirable that the implementation of such a network should  support coexistence of the quantum and classical signals in the same optical fiber transmission systems and share the same DWDM network components.

\section{IEQNET Architecture and Design} \label{sec:architecture}

\begin{figure}
    \centering
    \includegraphics[width=.7\textwidth]{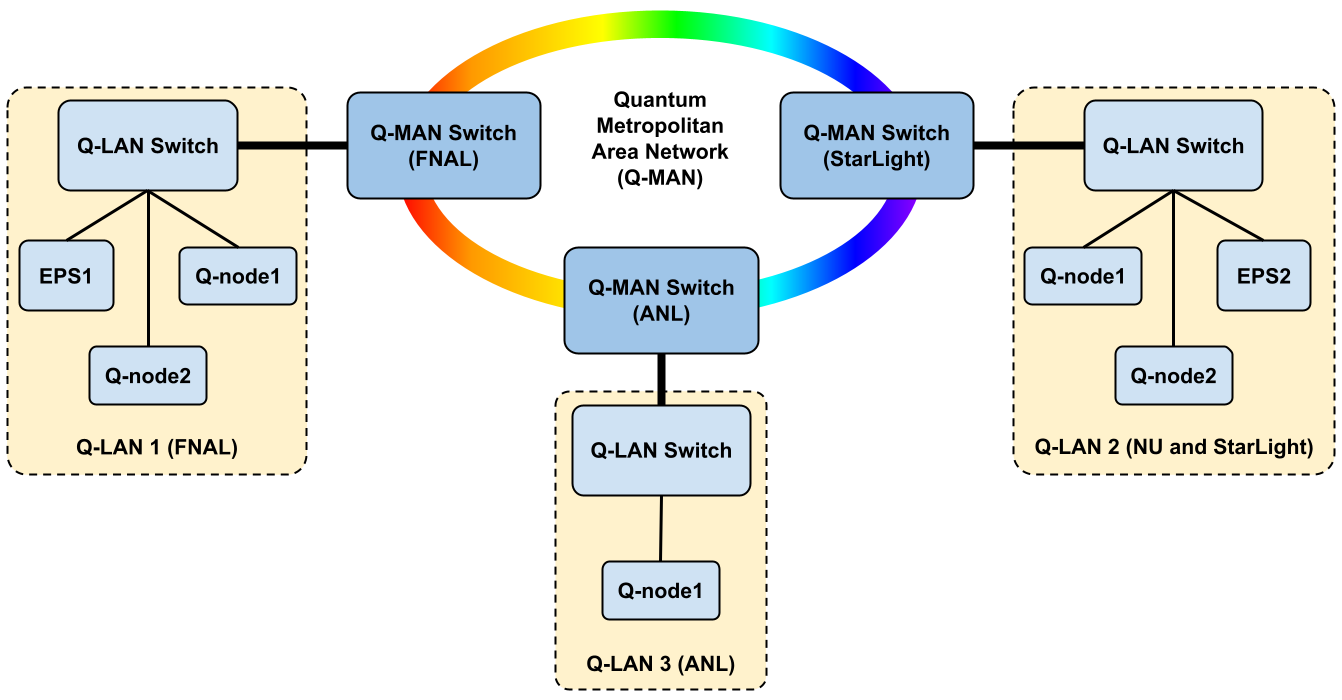}
    \caption{IEQNET topology}
    \label{fig:topo}
\end{figure}

\subsection{IEQNET Topology and Architecture} \label{sec:topo}
IEQNET consists of multiple sites that are geographically dispersed in the Chicago metropolitan area with sites at Northwestern University (NU), Fermi National Accelerator Laboratory (FNAL), Argonne National Laboratory (ANL), and a Chicago-based international communications exchange (StarLight). Each site has one or more quantum nodes (Q-Nodes), which can communicate data and generate and/or measure quantum signals. Q-Nodes are connected to SDN-enabled optical switches through optical fibers. The optical switches further connect with one another to form a meshed all-optical network. IEQNET contains three logically independent quantum local area networks (see Figure~\ref{fig:topo}): Q-LAN1 at FNAL, Q-LAN2 at NU and StarLight, and Q-LAN3 at ANL. The Q-LANs are connected by dedicated communication channels and additional dark fibers between FNAL and ANL (already operational), as well as between ANL and StarLight and FNAL and StarLight (these connections are part of the ESnet~\cite{esnet} plan for quantum network infrastructure deployment). A dark fiber is essentially an unused optical fiber cable with no service or traffic running on it. 

IEQNET has Q-Nodes that incorporate one or multiple entangled photon sources (EPSs) shared across the network, as well as Q-Nodes that incorporate Bell state measurement capabilities (BSM nodes), which are also connected to SDN-enabled optical switches through optical fibers. An EPS generates entangled photon pairs at N wavelengths, allowing a maximum of $N/2$ user-pairs to simultaneously share bipartite entangled photons. The N wavelengths from the EPS are typically distributed via N fibers to an all optical N$\times$N switch. Such switches are commercially available with low loss, high port counts, and transparency over the entire low-loss fiber wavelength range of 1270-1620 nm. These features allow efficient distribution of quantum photons to many users and open up wavelength options like the use of a 1310 nm band entangled photon source that allows for enhanced co-existence with classical data. A BSM node performs \textit{Bell state measurements}, projections of two qubit states onto Bell basis, and local qubit operations for incoming photon pairs.

\begin{figure*}[t]
\centering
  \includegraphics[width=\textwidth]{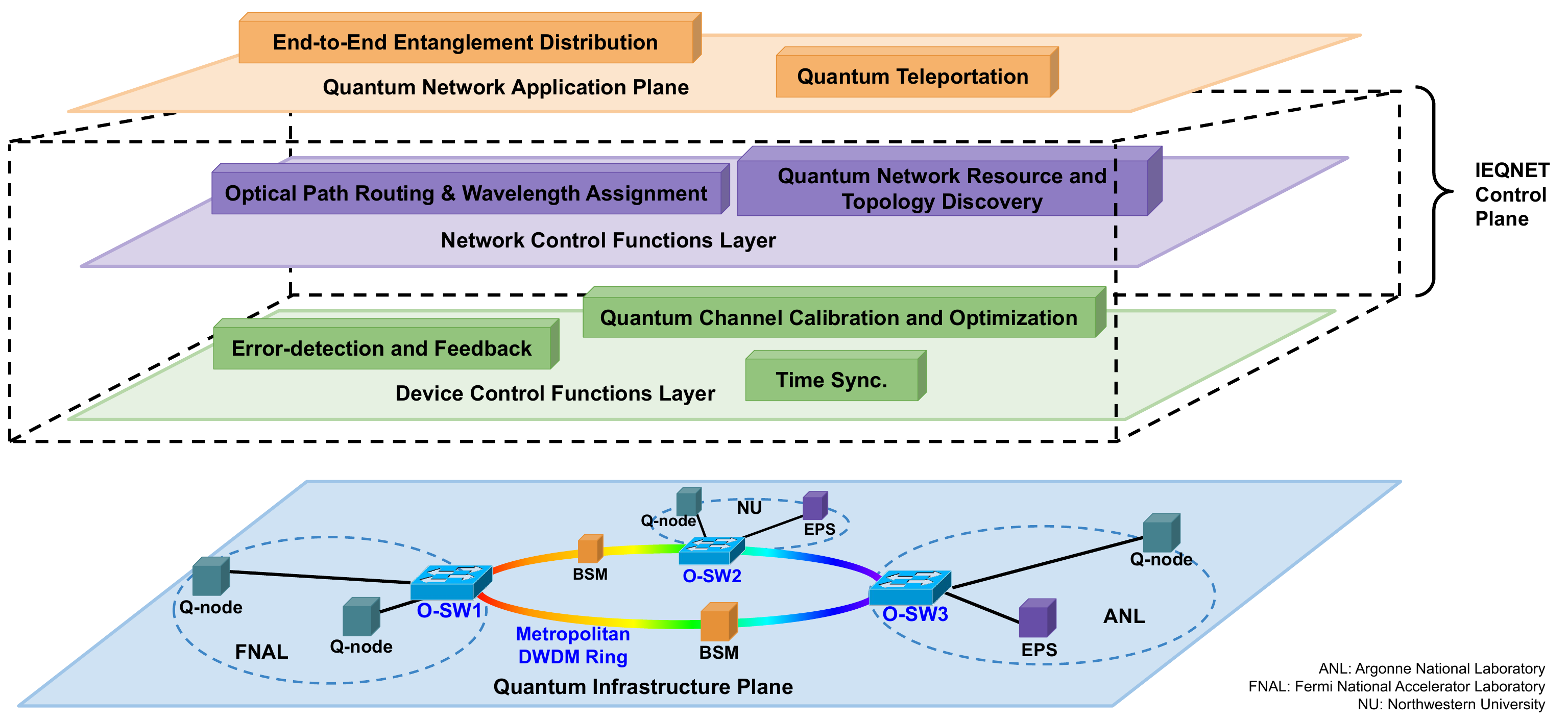}
  \caption{IEQNET's quantum networking architecture relies on three planes: infrastructure plane, control plane, and application. IEQNET's control plane is subdivided into Q-Node and Network control functions planes. See text for further description.}
  \label{fig:layerd-arch}
\end{figure*}

To satisfy the requirement of enabling entanglement distribution over a metropolitan area in the absence of quantum repeaters,
IEQNET implements a quantum networking architecture that resembles that of a classical circuit switching network. 
Repeaterless, metropolitan-scale quantum networks do not need the equivalent of classical packet switching in the quantum domain. These networks can operate under the circuit switching paradigm, that is, establishing a lightpath between Q-Nodes before communication starts. 
To achieve this, IEQNET's quantum networking architecture relies on three planes (see Figure~\ref{fig:layerd-arch}). 
\begin{itemize}
    \item \textit{Infrastructure Plane} deals with the physical connectivity of communicating quantum nodes. It is composed of all physical devices (e.g., Q-Nodes, EPS, and all-optical switches) and fiber optical links. It provides interfaces to allow the control plane (see next) to define parameters such as quantum channel frequencies, signal rates, and photon pulses used to represent quantum signals.
    \item \textit{Control Plane} deals with the classical control of individual devices and the network as a whole.
    \begin{itemize}
        \item \textit{Device control functions layer} consumes the interfaces provided by the infrastructure plane to ensure transmission of quantum signals and messages across quantum and classical channels. Among the functions of this layer we have error-detection and feedback, quantum channel calibration and optimization, and clock synchronization.
        \item \textit{Network control functions layer} performs wavelength routing and assignment in optical networks to establish lightpaths between Q-Nodes. This layer also performs quantum network resource and topology discovery. It also provides interfaces to the quantum network application plane to compose end-to-end services.
    \end{itemize}
    \item \textit{Quantum Network Application Plane} consumes interfaces provided by the control plane to compose end-to-end quantum networking services (e.g., end-to-end entanglement distribution and quantum teleportation) to users and applications.
\end{itemize}

Despite the current version of IEQNET being designed to support circuit switching style of quantum communication, the separation of planes allows for continuous evolution of the architecture.
For instance, once quantum memories and quantum error correction (QEC) are available, IEQNET can evolve into a third generation quantum repeated network~\cite{muralidharan2016optimal} by adopting a RuleSet-based control as described in~\cite{van-meter2022qi}.


\begin{figure}
\centering
  \includegraphics[width=.5\textwidth]{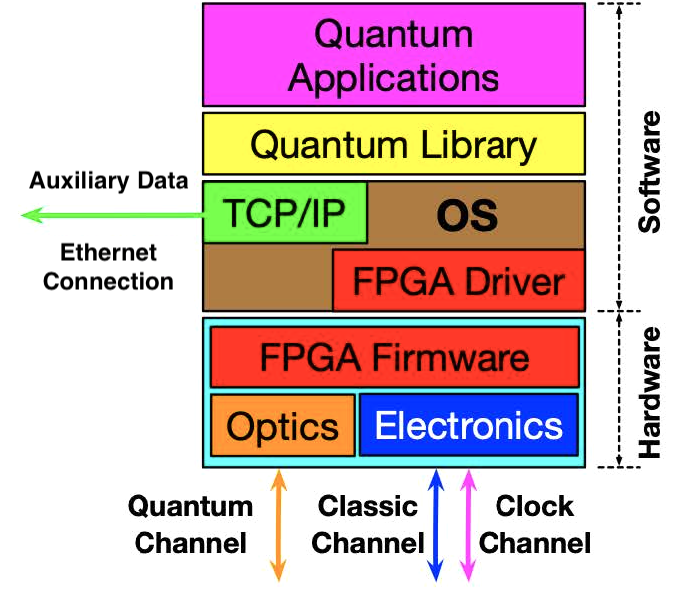}
  \caption{IEQNET Node implementation diagram. Each node performs conventional and quantum functions and has optical and electronic interfaces that are controlled by a FPGA with additional software layers running on top.}
  \label{Fig:Qnodes}
\end{figure}

\subsection{IEQNET Q-Node Design}

Much like their classical counterparts Q-Nodes in IEQNET, depicted in Fig. \ref{Fig:Qnodes}, represent the communication parties in a quantum network.
Every Q-Node performs both conventional (classical\footnote{We use ``classical'' and ``conventional'' interchangeably in this article, as a qualifier for classical network functions and/or signals}) and quantum functions.
The quantum functions that the node has to execute depend on the node type.
These range from single and entangled photonic qubit generation to qubit measurements and processing, including Bell-state measurements.
A Q-Node also performs conventional functions such as classical computation and communication. 
For example, a Q-Node will exchange messages with other Q-Nodes via conventional traffic channels to exchange the results of quantum measurements (e.g., to determine correlations or execute quantum protocols). 
A Q-Node is assigned an Internet Protocol (IP) address to uniquely identify the node. 
Meanwhile the endpoints of each conventional channel within the Q-Node are identified and addressed by a physical address (a classical device that contributes to the workflow of a classical function constitutes a ``conventional endpoint''). 
In addition, a Q-Node enumerates and numbers the quantum channels it connects to.
The \textit{\textless Q-Node IP, quantum channel \#\textgreater} pair is used to identify and address a particular quantum channel endpoint (a quantum device within the Q-Node that contributes to the workflow of a quantum function constitutes a ``quantum endpoint'').

Conventional channels serve three purposes in IEQNET: (i) as communication channels between Q-Nodes to exchange data and timing messages to enable quantum (correlation) measurements, (ii) embedded signals to allow calibration and stabilization functions (e.g., polarization stabilization, calibration of quantum basis measurements, or fiber-delay stabilization), (iii) as test channels for experimentation (e.g., to determine the limits of classical/quantum channel coexistence). 
We note that the calibration/stabilization could use quantum signals directly, but the use of larger magnitude classical signals when possible can provide enhanced functionality, such as reducing set-up time when the network changes the user connections or allows high-rate feedback control to compensate time-varying parameters such as polarization.

Routing is a fundamental network function. 
Multi-hop networks (those in which information has to travel across more than one networking devices) require a means of selecting paths through the network. 
Due to technology immaturity in quantum memory and quantum computation, IEQNET does not perform routing in the quantum domain, such as entanglement routing~\cite{pant2019routing}. 
Instead, SDN technology is used to perform traditional \textit{wavelength routing and assignment} in optical networks to establish lightpaths between Q-Nodes, or between Q-Nodes and entangled photon pair sources (EPSs), as appropriate (see Sec. \ref{sec:cp-design} for more details).

\subsection{IEQNET Control Plane Design} \label{sec:cp-design}
Mechanisms for generation, synchronization, and measurement of quantum states in networks are crucial for realizing a multitude of quantum information applications. They can allow high-quality distribution of entanglement throughout the network—provided errors are identified, their magnitude estimated, and steps taken for their correction. Orchestration and control mechanisms are especially important for performing advanced quantum communication tasks, such as quantum teleportation and entanglement swapping, which are based on quantum interference and thus are more susceptible to dynamical processes in a network environment, such as polarization rotations in the fiber channels or electronic control drift from local clock mismatches. IEQNET uses a centralized control approach in which SDN controllers monitor the status of key infrastructure plane metrics (e.g., loss on fiber links, status of optical switches, etc.). IEQNET's control and management software performs such functions as time synchronization, optical path routing and wavelength assignment for quantum and classical channels, channel calibration and optimization, and error detection and feedback.

\begin{itemize}
    \item \textit{Time Synchronization.} Synchronizing remote locations for distribution of entanglement and their use in subsequent applications is crucial for quantum networking. For the fiber channels, this is done by distributing clock pulses in the same fiber as the quantum signals, where permitted. 
    \item \textit{Routing} is a fundamental network function, and multihop all-optical networks require a means of selecting lightpaths through the network. IEQNET's underlying quantum network is a WDM-based all-optical network. 
    We use SDN technology to perform traditional routing and wavelength assignment (RWA)~\cite{zang2000review} to establish paths between Q-Nodes in IEQNET's quantum infrastructure plane. RWA is typically formulated as a multi-commodity ﬂow problem, an NP-hard problem that is typically solved with heuristic algorithms. For IEQNET, we represent the network by an undirected graph $G(V, E)$, where $V$ represents the set of nodes in the graph (Q-Nodes, BSM, EPSs, and optical switches) and $E$ represents the set of edges in the graph (fiber optical links). Each edge in the graph (i.e., fiber optical link in the network) has the following characteristics that contribute to the computation of the edge's metric/weight: link length, total number of wavelengths, number of wavelengths available, attenuation, etc. Each node in the graph may also contribute to the edge's metric/weight with attributes such as insertion loss, polarization-dependent loss (PDL), and polarization mode dispersion (PMD). An entanglement distribution request has a set of requirements $R$ such as minimum attenuation, maximum path length, minimum PLD, etc. We use the following shortest-path RWA (SP-RWA) algorithm (see Algorithm~\ref{algo:sp-rwa}) as the baseline to find a lightpath between two Q-Nodes given the network topology $G(V,E)$, source and destination Q-Nodes, a $k$ number of top paths, and the entanglement distribution request requirement $R$:
    \begin{enumerate}
        \item Find the set of $k$ shortest paths between the source-destination pair and sort the set according to the entanglement distribution requirements $R$
        \item If no path is found, the request is rejected
        \item For each path in the set, the wavelengths in the path are ordered according to the entanglement distribution requirements $R$
        \item For each wavelength in the path, verify that the wavelength is available for use
        \item The first available wavelength in the path that meets the entanglement distribution request requirement is assigned to the path and returned as the result
        \item If no available wavelength is found for any of the $k$ paths, the request is rejected
    \end{enumerate}

    \item \textit{Quantum Channel Calibration and Optimization.} The single-photon nature of quantum communication signals makes them extremely sensitive to noise on the quantum channels. In addition, as mentioned above, protocols such as teleportation require indistinguishability in spectral, temporal, spatial, and polarization properties of the two photons arriving at the BSM node. IEQNET employs several active and automated quantum-channel calibration and optimization mechanisms to minimize quantum-channel loss, reduce background noise, and compensate for polarization and delay drifts. 
    A common way to assess the indistinguishability is to perform a so called Hong-Ou-Mandel (HOM) experiment for the interfering photons by adjusting the relative time of arrival of the photons on the beam-splitter. This renders photons from completely distinguishable, corresponding to the case when the two photons arrive at different times and don't interfere with each other, to as indistinguishable as possible, corresponding to the case when two photons arrive at the same time at the beam-splitter. By increasing the HOM visibility we can ensure quantum indistinguishability. The monitoring of the HOM visibility provides feedback to compensate for the photons' relative time-of-flight, ensuring stable operation. Active polarization measurement and calibration using coexisting classical signals, such as clock pulses, is used to compensate for polarization drifts in fibers.
\end{itemize}

\begin{algorithm}
    \caption{Shortest-Path RWA (SP-RWA) algorithm}
    \label{algo:sp-rwa}
    \begin{algorithmic}[1]
        \State \textbf{Input:} $G(V,E)$, $src$, $dst$, $k$, $R$
        \Comment{$G(V,E)$ is a graph representing the network topology. $src$ and $dst$ are the source and destination Q-nodes, respectively. $k$ represents the number of paths and wavelengths to be found and $R$ represents the entanglement distribution requirements.}
        \State \textbf{Output:} $route$
        \Comment{Optimal lightpath between source and destination for entanglement distribution.}
        \State $route\gets$ \texttt{NULL} 
        \State $paths\gets$ \Call{FindAndSortPaths}{$G$, $src$, $dst$, $k$, $R$}
        \If{$paths$ is \texttt{NULL}}
            \State \Return \texttt{``Request Rejected''}
        \EndIf
        \For{each $path$ in $paths$}
            \State $wavelengths\gets$ \Call{SortWavelength}{$path$, $R$}
            \For{each $wl$ in $wavelengths$}
                \If{$wl$ is available}
                    \State $route\gets$ \Call{AssignWavelengthToPath}{$path$, $wl$}
                    \State \textbf{break}
                \EndIf 
            \EndFor
            \If{$route$ is not \texttt{NULL}}
                \State \Return $route$
            \EndIf
        \EndFor
        \State \Return \texttt{``Request Rejected''}
    \end{algorithmic}
\end{algorithm}

\subsection{IEQNET Control Plane Implementation} \label{sec:cp-impl}

\begin{figure*}[htb!]
    \centering
    \subfigure[]{\label{fig:cp-impl}\frame{\includegraphics[width=.4\textwidth]{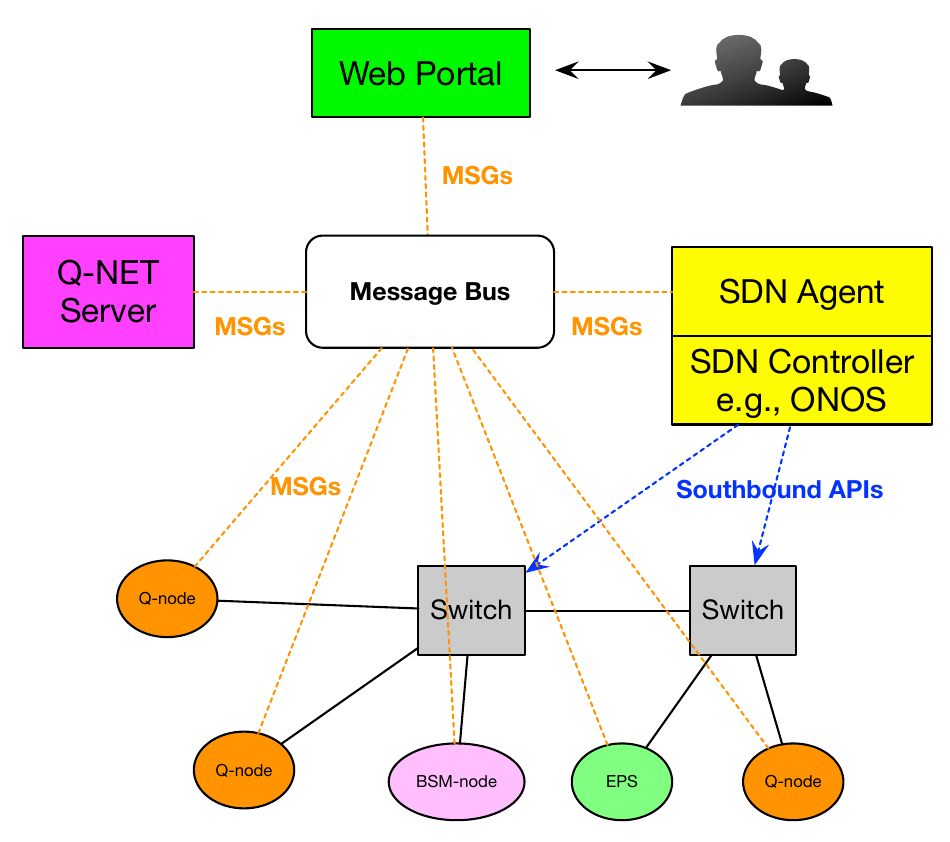}}}
    \subfigure[]{\label{fig:qnet-server}\frame{\includegraphics[width=.48\textwidth]{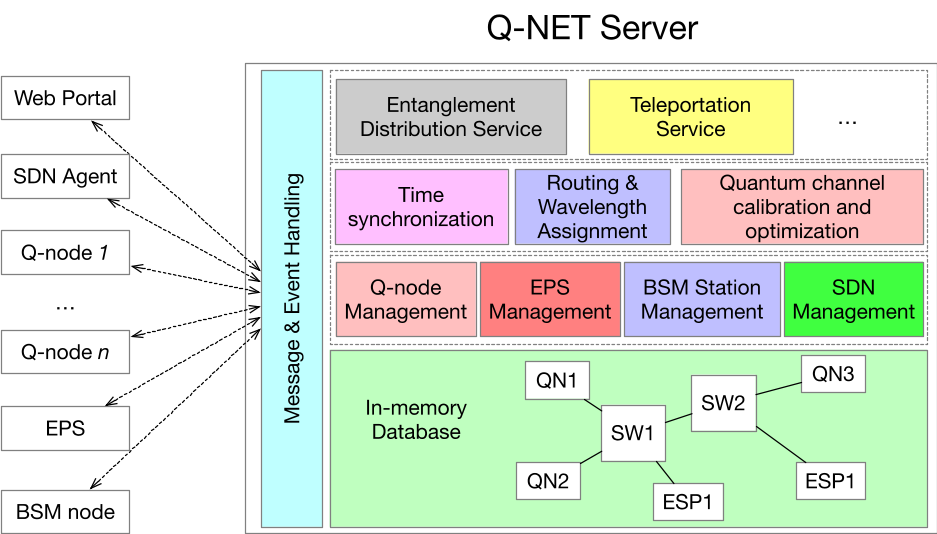}}}
	\caption{(a) IEQNET control plane implementation, composed by a centralized Q-NET Server, Web portal, and SDN agent all communicating via a message bus. (b) Q-NET Server software components including an in-memory database representing the network topology, management modules for Q-Nodes, EPS, BSM, and SDN agent, control functions such as time synchronization, RWA, and quantum channel calibration and optimization, quantum networking services such as entanglement distribution and teleportation; and a message \& event handling module.}
	\label{fig:ieqnet-cp}
\end{figure*}

We are implementing and deploying the SDN-based, logically centralized IEQNET control plane design described in section~\ref{sec:cp-design}, concurrently with the deployment of the IEQNET Q-Nodes and performing use-case demonstrators.  As illustrated in Figure~\ref{fig:cp-impl}, a logically centralized Q-NET server (magenta box in the figure) coordinates all activities in the network. This server manages and schedules various quantum network resources at the Q-Nodes (such as EPS, BSM, and quantum/classical channels) to perform key control and management functions of quantum networking services (e.g., entanglement distribution and quantum teleportation for the IEQNET use cases).
Figure~\ref{fig:qnet-server} shows the software components of the Q-NET server that include an in-memory database representing the network topology; management modules for Q-Nodes, EPS, BSM, and SDN agent; control functions such as time synchronization, RWA, and quantum channel calibration and optimization; and quantum networking services such as entanglement distribution and teleportation. The Q-NET server communicates with the rest of IEQNET control plane components through the Message and Event Handling module.

A web portal (see top green box in Figure~\ref{fig:cp-impl}) authenticates, authorizes, and audits users and applications, and allows them to access IEQNET services and functions. For example, for an entanglement distribution service request, the following information is conveyed to the Q-NET server via the web portal: the credentials of the task submitter, the Q-Nodes involved, and the entanglement distribution requirements, such as qubit type, rate, duration, etc. The Q-NET server uses this information to schedule and broker resources for the task. In addition, users are able to browse the quantum network topology or monitor the system/site status via the web portal.

As the network expands and/or changes with nodes or connections joining or going offline, an SDN agent keeps track of the quantum network topology and traffic status with the aid of SDN controllers (see yellow boxes in Figure~\ref{fig:cp-impl}). The SDN agent is also responsible for reliably updating SDN-enabled switch rules, as requested by the Q-NET server, to assign paths for quantum, clock, and classical signals/messages. SDN controllers are open-source network operating systems, such as the Open Network Operating System (ONOS)~\cite{onos}. The SDN agent accesses SDN controllers through northbound application programming interface (APIs). A northbound API allows other applications to send commands to an SDN controller. One or multiple SDN agents will be deployed, depending on the size of the quantum network.

In this implementation, the Q-NET server communicates with other entities through a message queuing telemetry transport (MQTT) based message bus (see white box and orange lines in Figure~\ref{fig:cp-impl}). Such a control plane design offers flexibility, robustness, and scalability.

\subsection{IEQNET Classical Control Protocol Suite} \label{sec:protocol}
In this section we describe the IEQNET classical control protocol suite, which is composed of two main protocols: quantum network resource and topology discovery protocol 
and the protocol for handling entanglement distribution requests. 
The classical control protocol suite runs inside the Q-NET server of IEQNET's control plane implementation.
Figures \ref{fig:discovery} and \ref{fig:res-mgmt} in the Appendix show sequence diagrams of the proposed protocols.
As both these protocols remain in the control plane of IEQNET, their efficiency can only be evaluated with classical metrics. An initial qualitative evaluation of these protocols can be inferred by simply looking at the efficiency of SDN control interfaces and MQTT buses. We expect that IEQNET control protocols would be able to converge to solutions in the order of tens of milliseconds for a metropolitan quantum network of three sites using the proposed protocols.

The discovery protocol (see pseudocode in Algorithm~\ref{algo:qnrdp}) starts when each quantum networking resource (e.g., Q-Nodes, EPS, BSM-nodes, and switches) load their own configuration from file at initialization. The next step is for the SDN agent to discover the network topology through the SDN controller's southbound API. A southbound API allows an SDN controller to communicate with networking devices such as all-optical switches. As all-optical switches are passive devices and thus classical active mechanism for topology discovery will not work, we have extended ONOS's link discovery service to build the topology from information loaded in a tag field on each optical port's configuration. Quantum network resources register to the Q-NET server by sending their features and connectivity information. The Q-NET server will request the topology from the SDN agent and will subsequently ask the SDN agent to verify the connectivity information provided by individual quantum resources. Once the topology has been verified, the Q-NET server will build a topology graph based on the updated topology and present it to users through the Web portal. 
This cross-verification is necessary because the current prototype of the IEQNET controller does not support active discovery (as already mentioned). As we work with configuration files, the only way the SDN Agent can know a Q-Node is connected to its designated port is by receiving a message from the Q-NET server after the Q-Node registered.
This discovery protocol keeps running as quantum network resources can come and go. Furthermore, the SDN agent has the capability to notify the Q-NET server of topology changes asynchronously.

\begin{algorithm}
    \caption{Quantum Network Resource and Topology Discovery Protocol}
    \label{algo:qnrdp}
    \begin{algorithmic}[1]
        \State \textbf{@ all Q-Nodes} 
        \State \Call{Initialize}{config\_file}
        \State \textbf{@ SDN Agent}
        \State $topo\gets$ \texttt{NULL}
        \For{each $switch$ in $switches$}
            \State $neighbors\gets$ \Call{QueryNeighbors}{$switch$}
            \State $topo\gets$ \Call{UpdateTopo}{$switch$, $neighbors$}
        \EndFor
        \State \textbf{@ all Q-Nodes} \Comment{in parallel} 
        \State \Call{SendReg}{$q\_node\_info$, QNETServer\_address}
        \State \textbf{@ Q-NET Server}
        \While{\texttt{RegistrationEvent}}
            \State $quantum\_resources\gets$ \Call{UpdateQResources}{$q\_node\_info$}
        \EndWhile
        \State \Call{QueryTopo}{SDNAgent\_address}
        \State \textbf{@ SDN Agent}
        \State \Call{SendTopo}{$topo$, QNETServer\_address}
        \State \textbf{@ Q-NET Server}
        \For{each $q\_node$ in $quantum\_resource$}
            \State $topo\gets$ \Call{VerifyConnectivity}{$q\_node\_info$, SDNAgent\_address}
        \EndFor
        \State \Call{SendTopo}{$topo$, WebServer\_address}
        \State \textbf{@ Web Server}
        \State \Call{BuildTopoVisualization}{$topo$}
    \end{algorithmic}
\end{algorithm}

The entanglement distribution protocol (see pseudocode in Algorithm~\ref{algo:ent-dist-proto}) starts with a user requesting entanglement distribution between Q-Node1 and Q-Node2. The Q-NET server will analyze this request and choose an EPS that meets the requirements specified by the user. Upon acceptance of the request, the Q-NET server will execute path routing and wavelength assignment and will establish the paths among involved entities via the SDN agent. Q-NET server will notify Q-Node1 and Q-Node2 when paths are established and initiate path verification, which involves a series of active probes from EPS to Q-Nodes (and vice versa) using both classical and quantum light. After path verification, the Q-NET server will initiate calibration and optimization processes (as described in Section~\ref{sec:cp-design}) for the requested service. Once all entities send the READY signal to the Q-NET server, the entanglement distribution process starts. Q-Nodes will collect measurements until they have a long enough ebit string for the upper layer application. At that point they will send the END signal to the Q-NET server to stop entanglement distribution. Periodically during entanglement distribution, the Q-NET server will re-initiate the calibration and optimization processes. After the Q-NET server stops the EPS, all measurements will be stored at the Q-NET server and the user will be able to access them through the Web portal.

\begin{algorithm}
    \caption{Entanglement Distribution Protocol}
    \label{algo:ent-dist-proto}
    \begin{algorithmic}[1]
        \State \textbf{@ User} 
        \State \Call{UserReq}{$ent\_dist\_reqs$, $qubit\_type$, $start\_time$, $end\_time$, $[QNode1, QNode2]$, $measurement\_basis$}
        \State \textbf{@ Q-NET Server}
        \State $req\gets$ \Call{ReceiveReq}{$UserReq$}
        \State $eps\gets$ \Call{SelectEPS}{req}
        \If{$eps$ is \texttt{NULL}}
            \State \Call{RejectUserRequest}{}
        \Else 
            \State $R\gets$ \Call{ComputeRequirement}{$ent\_dist\_reqs$, $qubit\_type$}
            \State $path1\gets$ \Call{SP-RWA}{$topo$, $QNode1$, $eps$, $k$, $R$}
            \State $path2\gets$ \Call{SP-RWA}{$topo$, $QNode2$, $eps$, $k$, $R$}
            \State \Call{SetupPath}{$path1$, SDNAgent\_address}
            \State \Call{SetupPath}{$path2$, SDNAgent\_address}
            \State \Call{SendNotification}{\texttt{``path ready''}, QNode1\_address}
            \State \Call{SendNotification}{\texttt{``path ready''}, QNode2\_address}
            \State $path1\_ready \gets False$
            \State $path2\_ready \gets False$
            \While{!$path1\_ready$ and !$path2\_ready$}
                \State $path1\_ready \gets$ \Call{VerifyPath}{$path1$}
                \State $path2\_ready \gets$ \Call{VerifyPath}{$path2$}
            \EndWhile
            \State $esp\_status \gets$ \Call{CalibrationAndOpt}{EPS\_address, $R$}
            \State $qnode1\_status \gets$ \Call{CalibrationAndOpt}{QNode1\_address, $R$}
            \State $qnode2\_status \gets$ \Call{CalibrationAndOpt}{QNode2\_address, $R$}
            \If{$esp\_status$ == \texttt{READY} and $qnode1\_status$ == \texttt{READY} and $qnode2\_status$ == \texttt{READY}}
                \State \Call{SendNotification}{\texttt{``collect measurements''}, QNode1\_address}
                \State \Call{SendNotification}{\texttt{``collect measurement''}, QNode2\_address}
                \State \Call{SendNotification}{\texttt{``start entanglement distribution''}, EPS\_address}
            \EndIf
        \EndIf
        \State \textbf{@ EPS}
        \State \Call{StartEntDist}{$ent\_dist\_reqs$, $qubit\_type$}
        \State \textbf{@ Q-Node1 and Q-Node2} \Comment{in parallel}
        \State \Call{CollectMeasurements}{$ent\_dist\_reqs$, $qubit\_type$}
        \If{$ebit\_string \geq LENGTH$}
            \State \Call{SendNotification}{\texttt{``END''}, QNETServer\_address}
        \EndIf
    \end{algorithmic}
\end{algorithm}
\section{IEQNET Demonstrators} \label{sec:evaluation}
In this section we present experimental results that demonstrate system capabilities necessary for the deployment of the IEQNET quantum network architecture. These involve quantum teleportation and quantum/classical signal co-existence experiments performed at both Q-LAN1 (Fermilab) and Q-LAN2 (Northwestern University) of IEQNET. Demonstrating co-existence in real-world conditions is a major goal for IEQNET and a very desirable development for the deployment of quantum networks. Classical communications are a fundamental part of many quantum communications protocols that require the transmission of the results of quantum measurements amongst the node connections, such as remote state preparation and  teleportation. Furthermore, classical signals for time synchronization in quantum networks will be required. Beyond purely quantum-based applications, future quantum networks will likely have fiber connections that are already populated with light from the classical internet, which is predominantly in the C-band. If quantum communications can coexist in the same fibers carrying such high powers, the entire classical internet fiber infrastructure is available for quantum network deployment. This would significantly reduce the cost of deploying quantum networks as dark fibers will not be required and increase the number of available fiber connections between node locations. Dark fibers are expensive to lease due to the high demand for installed fibers from the expansion of the classical internet. We demonstrate engineering approaches to achieve this such that the network could be deployed anywhere within the fiber infrastructure. 

Q-LAN1 has nodes deployed in two separate locations at Fermilab, connected by $\sim$2.5 km of optical fiber. Q-LAN1 is used to demonstrate time-bin qubit teleportation, although these initial tests are performed in a single location to allow for experimental flexibility and enable faster debugging. We perform teleportation over 44 km of lab-deployed fibers at 1536 nm and achieve above 90~$\%$
teleportation fidelities with a semi-autonomous system that can sustain stable operation via modern data-acquisition systems and integrated feedback mechanisms. We also discuss the commissioning results of a clock distribution system, co-existing in the same optical fiber as the quantum channel, and where we measured a time jitter of 5~ps. These co-existing clock distribution systems will pave the way towards synchronizing multiple nodes in remote locations and thus allowing more complex quantum protocols.

Q-LAN2 connects the Quantum Communications Lab at Northwestern University in Evanston, IL to the Starlight Communications Facility located on the Chicago campus of Northwestern University. This underground fiber link is 22.8 km, with the option of operating in loop-back mode to reach 45.6 km. The experimental measurements reported in this section operate at the Northwestern site using the 45.6 km loop-back configuration. The Starlight facility also has access to a short-reach fiber connection to 600 S. Federal building and a 16$\times$16 all-optical switch. Q-LAN2 distributes polarization entangled photons, and includes the use of classical alignment signals to account for the unknown polarization transfer function of the distribution fiber. This network uses the O-band (1310 nm band) for the quantum signal, which is chosen so that substantial amounts of co-existing classical communications in the C-band (1550 nm band) can be tolerated on a shared optical fiber. Our experiments and data analysis are guided and supported by a phenomenological model which can be quickly compared with experimental data. We discuss the model and its utilization in our results.

\subsection{Quantum Teleportation of Time-bin Qubits and Coexistence with Telecommunication O-band Classical Signals at Q-LAN1} \label{sec:teleportation}
The Q-LAN1 nodes are deployed in two separate locations at Fermilab: one at the D0 Assembly Building (DAB) and the other at the Fermilab Computing Center (FCC). The two labs supporting the Q-LAN1 nodes at these two locations are connected by $\sim$2.5 km of optical fiber going through a 16$\times$16 (Polatis 6000s Ultra) all-optical switch.
The Q-LAN1 setup with its two  locations, short separation by field-deployed fiber, and optical switch serves as an internal IEQNET ``testbed'' to test and validate performance of IEQNET protocols before connecting the three Q-LANs (first Q-LAN1 and Q-LAN3 at Argonne, and then Q-LAN2, the sequence in the order of increasing distance).  The first set of Q-LAN1 experiments were performed at the DAB lab.  

\subsubsection{Quantum Teleportation}

\begin{figure}
    \centering
    \includegraphics[width=.9\textwidth]{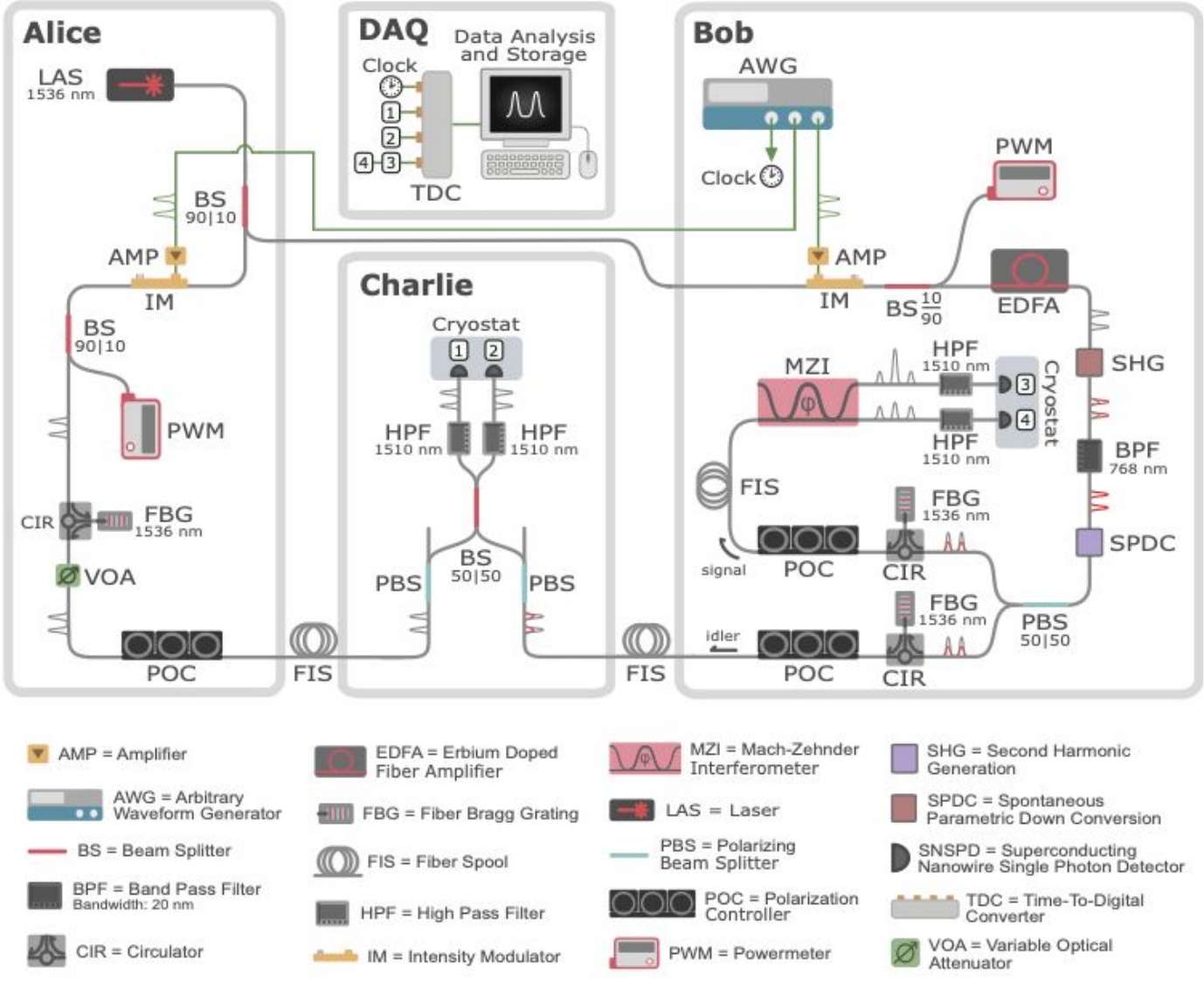}
    \caption{Schematic diagram of the quantum teleportation system consisting of Alice, Bob, Charlie, and the data acquisition (DAQ) subsystems. One cryostat is used to house all SNSPDs, it is drawn as two for ease of explanation. Detection signals generated by each of the SNSPDs are labelled 1-4 and collected at the TDC, with 3 and 4 being time-multiplexed. All individual components are labeled in the legend, with single-mode optical fibers (electronic cables) in grey (green), and with uni- and bi-chromatic (i.e., unfiltered) optical pulses indicated.}
    \label{fig:figure5}
\end{figure}

\begin{figure*}[htb!]
    \centering
    \subfigure[]{\label{fig:figure6a}\includegraphics[width=.48\textwidth, trim=0 0 0 0.75cm, clip]{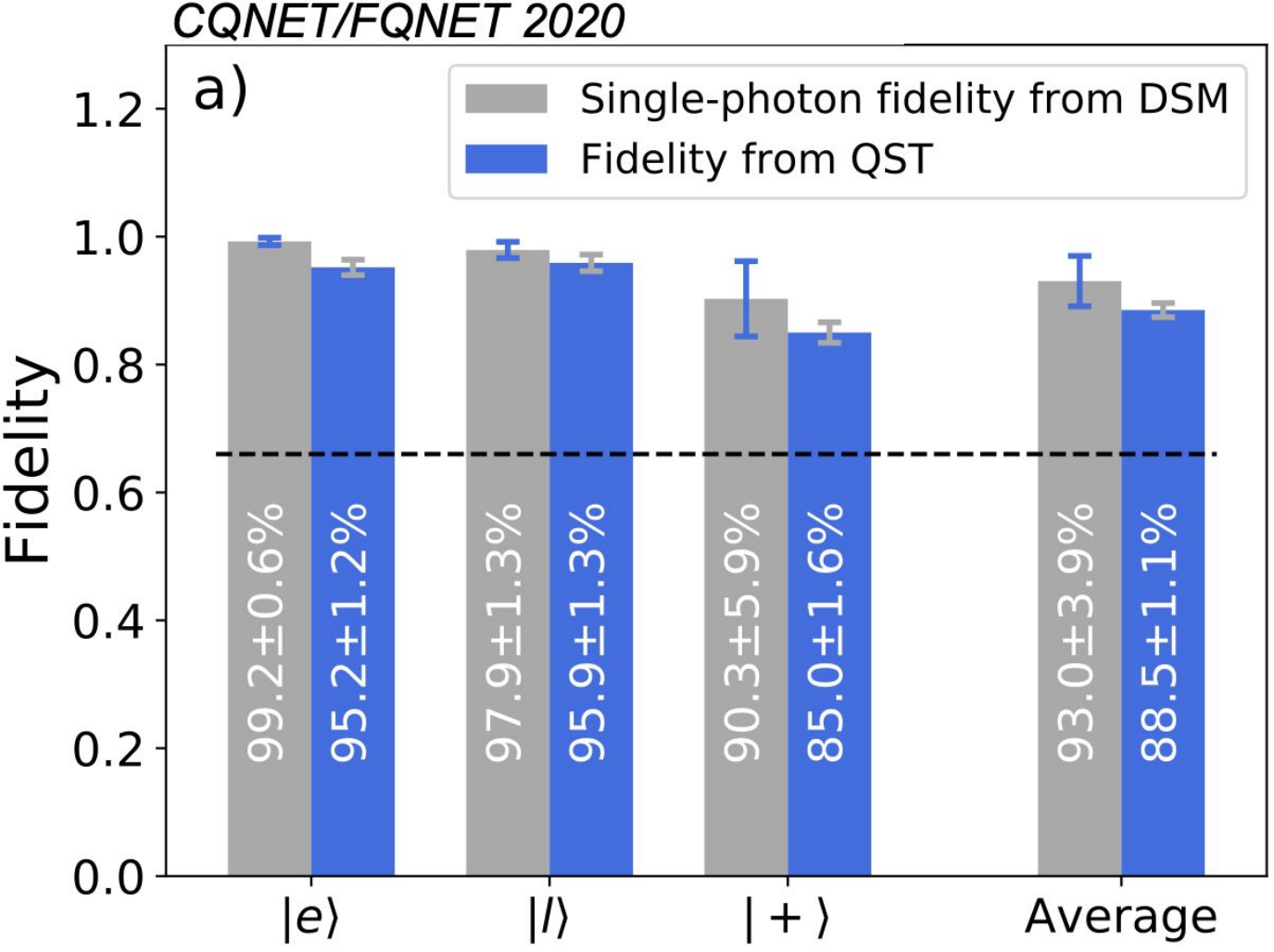}}
    \subfigure[]{\label{fig:figure6b}\includegraphics[width=.46\textwidth]{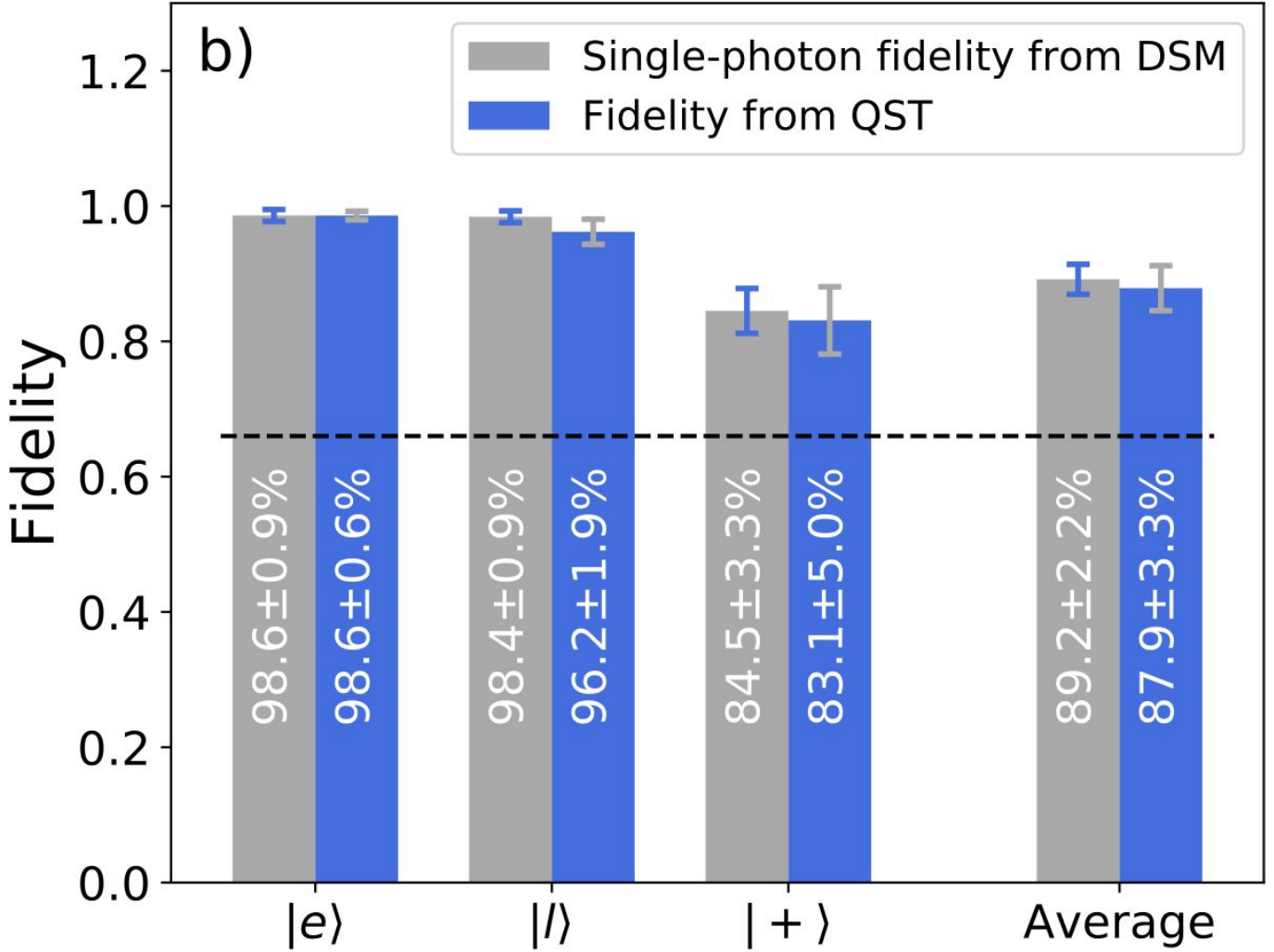}}
	\caption{Quantum teleportation fidelities for $|e>_A$, $|l>_A$, and $|+>_A$, including the average fidelity. The dashed line represents the classical bound. Fidelities using quantum state tomography (QST) are shown using blue bars while the minimum fidelities for qubits prepared using $|n=1>$, $F^d_e$, $F^d_l$, and $F^d_+$, including the associated average fidelity $F^d_avg$, respectively, using a decoy state method (DSM) is shown in grey. Panels a) and b) depict the results without and with additional fiber, respectively. Uncertainties are calculated using Monte-Carlo simulations with Poissonian statistics}
	\label{fig:figure6}
\end{figure*}

The Q-LAN1 fiber-based experimental system, summarized in the diagram of Figure~\ref{fig:figure5}, allows us to demonstrate a quantum teleportation protocol in which a photonic qubit (provided by Alice) is interfered with one member of an entangled photon-pair (from Bob) and projected (by Charlie) onto a Bell-state whereby the state of Alice's qubit can be transferred to the remaining member of Bob's entangled photon pair. Up to 22 (11) km of single mode fiber is introduced between Alice and Charlie (Bob and Charlie), as well as up to another 11~km at Bob, depending on the experiment. All qubits are generated at a clock rate of 90~MHz, with all of their measurements collected using a data acquisition (DAQ) system. The measured teleportation fidelities with and without the additional fiber are presented in Figure~\ref{fig:figure6}. The rates of the teleportation in the experiment are dependent on mean photon number of Alice's source. For a mean photon number of $\sim$0.01, the average rate of successful teleportation is $\sim$4 Hz without the fiber spools and $\sim$0.1 Hz with the full fiber. To illustrate network compatibility, teleportation is 
facilitated using semi-autonomous control, monitoring, and synchronization systems, with results collected using scalable acquisition hardware. Our system can be run remotely for several days without interruption.
The Q-LAN1 systems are identical to those utilized in~\cite{valivarthi2020teleportation}, with the DAB lab nodes being part of these original experiments. Our qubits are also compatible with erbium-doped crystals, e.g. Er:Y2SiO5, that is used to develop quantum network devices like memories and transducers~\cite{miyazono2016coupling,welinski2019electron,lauritzen2010telecommunication}. 
The collaboration is currently working towards achieving teleportation rates appropriate for practical, real-world applications by increasing the clock rate from 90~MHz to 4~GHz. The improvement of rates in conjunction with integration of quantum memories will be crucial for the development of quantum repeaters.

\subsubsection{Clock Distribution}
\begin{figure}
    \centering
    \includegraphics[width=.5\textwidth, trim=0 0 0 0.75cm, clip]{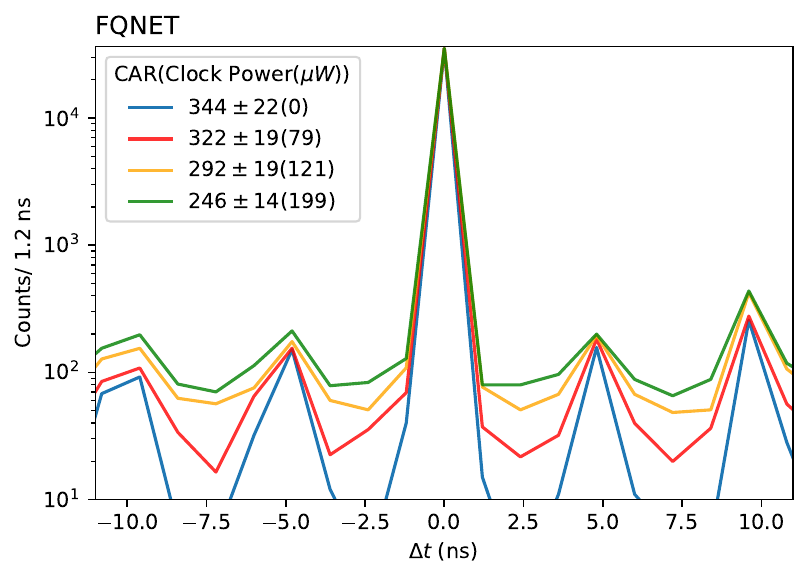}
    \caption{Time difference between photons from the pair source ($\Delta t$). The main peak at around $\Delta t = 0$ corresponds to the coincidence peak while the secondary peaks are the accidental counts. Different input power for the co-propagating clock signal in the O-band (1310~nm) are shown in different colors. The increasing heights of the accidental peaks with the higher clock power indicate more Raman noise.}
    \label{fig:car}
\end{figure}

While quantum teleportation has been achieved with state-of-the-art fidelities and overall performance parameters capable of supporting network functions, the demonstrations has nevertheless been restricted to a single node. In a real world network containing more than one node, clock distribution to synchronize the various nodes is critical. Since photons are identified by recording their times of generation and detection, an accurate clock distribution system must be in place to reduce the error rate to a negligible level. To this end, we developed a clock distribution system based on the demonstration presented in~\cite{sync_paper}. The system relies on a stable clock oscillator measured to have a jitter below 700~fs. The oscillator is used to synchronized all the RF components of the experiment in the main Q-node and additionally drive a 200~MHz clock signal in the O-band which is directed to two independent nodes to adjust their local clocks. C-band photon pairs, originating from a source based on spontaneous parametric down-conversion, are co-transmitted with the classical clock signal. We find that the clock distribution system can allow for high-fidelity qubit distribution despite the presence of (Raman) noise. We measure the coincidence to accidental ratio (CAR) of the photons arriving at the remote Q-node. Figure~\ref{fig:car} shows the difference in arrival time from the pair source ($\Delta t$) for different input clock powers. We measure that the clock distribution reduces the CAR from $344\pm22$ to $246\pm14$, this resulting CAR will ensure high-fidelity qubit distribution using this clock system. The clock distribution was measured to have a jitter below 5~ps. 
The observed timing jitter between clocks at the central and end nodes, suggests that our method can be used for high-rate networks. The deployment of such a system will pave the road towards implementing important multi-node quantum functions for scalable networks.

\subsubsection{Entanglement Swapping}
Furthermore, by building on the current capabilities, the systems are being upgraded towards entanglement swapping~\cite{zukowski1993event}, a key requirement for building long distance quantum networks. By performing a BSM between individual members of two entangled photon sources, entanglement is swapped onto the photons that have never interacted before.  Using the newly commissioned FCC Q-node, we have measured the indistinguishability of the photons from two photon pairs sources using a Hong-Ou-Mandel analysis~\cite{hong1987measurement}. The experimental results closely match a model of our experimental setup which predicts, under these experimental conditions, a high swapping fidelity.

\subsection{Polarization Entanglement Distribution with Coexisting C-band Classical Light over Real-World Installed Fiber at Q-LAN2} \label{sec:coexistence}

In addition to time-bin entanglement distribution, IEQNET also supports the distribution of polarization entanglement. Here, we demonstrate methods for monitoring and compensating polarization birefringence effects during transmission of photon pairs over installed fiber, which is required for quantum applications and measurements in polarization entanglement-based networks. Classical alignment signals are used to transmit information about the birefringent rotations during transmission over a fiber connection to polarization control systems (PCSs) at receiving nodes. Using this information, inverse unitary transformations are applied such that each Q-node that is connected to a source at $\text{Q-node}_A$ shares a universal polarization reference frame defined during the entanglement generation at the source. 

Further, the Q-LAN2 installed fiber connection explores the implementation of alternative wavelength allocation for quantum signals to make the physical layer more robust to noise photons generated from classical communications coexisting inside the same fiber connection~\cite{thomas_entanglement_2022}. The polarization entangled quantum signal is allocated to the O-band, which is shown to be optimal for scenarios in which the fiber connection is occupied with C-band classical communication with high enough power that the amount of generated photon noise is fatal to C-band quantum communication~\cite{chapuran2009optical}. 


\subsubsection{Coexistence with C-band Classical Light}


The wavelength allocation of quantum and classical signals to implement quantum network protocols is flexible when dark fibers are available, allowing the lowest-loss C-band to be used for optimal rate-loss quantum signal transmission. In this scenario, classical signals for quantum network control can be allocated to other wavelength bands (such as the L or O-bands) to minimize noise cross-talk into the quantum bands. However, the design of fully integrated networks must presume that high-power classical communications in the same fiber will unavoidably occupy the C-band, making quantum signal allocation to the C-band impractical at some threshold power level due to the presence of spontaneous Raman scattering (SRS), four-wave mixing (FWM), and amplified spontaneous emission (ASE) from optical amplifiers. This noise disrupts the ability to perform quantum communications protocols in the C-band, requiring the reconsideration of the quantum physical layer's wavelength allocation scheme. We mitigate the noise limitations caused by C-band classical coexistence by wavelength allocating the quantum entangled photon pairs to the O-band. Due to the comparatively narrow bandwidths of FWM and appropriate filtering of the ASE, these effects have no noticeable cross-talk into the O-band. Further, since the O-band is far detuned from the C-band on the anti-Stokes side ($\sim$ 35 THz), noise photons generated from SRS are significantly reduced. Since the O-band has reasonably low loss in standard installed fibers ($\sim$ 0.33 dB/km), metro-scale quantum communication can still be performed while noise from classical communications is drastically reduced compared to the C-band. 

There have been numerous experimental implementations of a variety of coexistence scenarios in the context of quantum key distribution (QKD) using weak coherent states (WCS) from attenuated laser light with a strong focus on the O-band/C-band quantum/classical wavelength allocation scheme described above~\cite{ eraerds_quantum_2010, Yuan:19, wang_experimental_2015, aleksic_perspectives_2015, wang_long-distance_2017, mao_integrating_2018, geng_coexistence_2021}. These studies were primarily performed via single quantum channel schemes with WCS. However, fully operational quantum networks beyond QKD will require the distribution of quantum entanglement. This means that two channels must be considered simultaneously, one for each photon in the pair of the entangled state. In contrast to the single quantum channel experiments, the signal to noise trade-off for entanglement-based networks depends on the coincidence detection of correlations in the signal and idler pair's time, frequency, and entangled degree of freedom. Due to the correlations in time-of-arrival and energy conserving frequency correlations, tight temporal and spectral filtering in coincidence detection can be used to distinguish the quantum entangled signals from the uncorrelated background noise generated from classical communications. 

As a demonstration of the above-mentioned engineering considerations, we show experimentally that metropolitan scale entanglement-based quantum communications can be achieved in installed fiber using the O-band with copropagating classical C-band power levels that would make C-band quantum transmission infeasible. We show that polarization-entangled quantum light can successfully copropagate with milliwatt power-level classical light over 45.6 km of installed underground fiber. One photon from a polarization entangled source at the Northwestern Evanston location is transmitted over the Q-LAN2 connection to the Starlight facility where it then loops back to the Evanston lab, while the other photon is kept locally. Polarization analyzers and low-dark-count superconducting nanowire single photon detectors (SNSPDs) are used for quantum characterization of the returning noise degraded quantum signal, which have ~30\% detection efficiency in the O-band. The equivalent measured loss rate in the underground fiber link is 0.43 dB/km at 1310 nm, which is higher than expected in modern fibers typically used in laboratory experiments, making the equivalent loss closer to a 60 km distance if newer fibers and low loss 
splicing were used. The use of O-band/C-band quantum/classical wavelength allocation as well as tight temporal and spectral filtering at receiver nodes allows us to achieve high copropagating classical powers (about 7 dBm) while still maintaining nonclassical visibility (> 70.7\%) in polarization entanglement two-photon interference.

\subsubsection{Polarization Entanglement Distribution Calibration}

The quantum networking layer will control routing and assignment of lightpaths between nodes, which means every time a new switched fiber connection has been established, re-calibration of the pair of paths needs to be performed. Further, occasional re-calibration will also be required. Inevitably, time-dependent birefringence due to a variety of environmental disturbances (temperature fluctuations, fiber stresses on hanging fiber, etc.) will cause an initial polarization state to drift over time. Thus, a crucial part of polarization entanglement network control is to ensure that drift from the intended quantum state can be accurately and reliably monitored to re-establish the desired quantum correlations between node pairs. Here we explain a simple protocol for polarization calibration between nodes in the network using broadband classical light sources that are built into the entanglement source's design.

Basis-alignment can be performed via the transmission of alignment signals which can be analyzed to extract the polarization transformations that occur from the source to receivers. Then, unitary operations at receiving nodes can be performed to compensate for birefringence in the fibers and establish well-defined relative polarization reference frames amongst nodes in the network for performing correlated measurements. The design of polarization monitoring highly depends on the timescales of the drift from the intended transmitted state. For slowly varying fiber birefringence, alignment can be maintained for hours of operation with little drift from the intended quantum state, however some fiber connections may have drift timescales where re-calibration may be required more frequently. In some situations, particularly when loss is less prevalent and drift timescales are slow, using the output of the quantum entanglement source itself can be used to align measurement basis. This method is limited by the transmission rate of the quantum source, which significantly increases the calibration duration and complexity due to low singles or coincidence count rates, especially over long-distance fiber connections with high loss. Active monitoring of polarization rotations can be performed by multiplexing in low power classical light at neighboring wavelengths~\cite{xavier_full_2008, xavier_experimental_2009} at the expense of increased design complexity and coexisting classical channels, which may introduce some noise into the quantum band.

We choose to use a broadband classical alignment signal that is built into the O-band quantum entanglement source. The broadband nature of the signal means that when the alignment signal is turned on, classical light is then sent throughout the network to each node connection deriving from the EPS for basis alignment. Since the alignment signal is in the same wavelength band as the quantum signal, it contains wavelength dependent polarization effects and propagates the fiber network connections defined by the wavelength routing/assignment and switching connections for quantum channel configurations without additional modifications to the network physical design. This introduces light into the quantum band, temporarily disrupting quantum communications. However, the classical nature of the alignment signal allows us to de-couple the alignment procedure from quantum source rates, allowing the signal to overcome long distance or high loss fiber connections by arbitrarily controlling the power transmitted and thus significantly reducing off-line calibration time due to the strong feedback signal for alignment regardless of link distance. For situations with many connections, it may be desirable to only select a few channels that need alignment. By including WDMs and variable attenuators, our signal can be 
modified to selectively send to a single or subset of receiving nodes such that quantum communications are not disrupted over the entirety of the network connections. 

Our alignment procedure consists of multiplexing in broadband amplified spontaneous emission (ASE) covering the entire O-band to transmit a signal $|V \ align\rangle$ for alignment in the H/V basis followed by a $|diag \ align\rangle$ signal used to align in the D/AD basis via polarization control at each node. The two non-orthogonal signals allow for full alignment of the Poincare sphere at nodes Bob and Charlie to that defined during entanglement generation at Alice \cite{xavier_full_2008}. The broadband spectrum of the ASE is carved by the WDM filters defined for transmission of the entangled pairs, which allows for wavelength dependent birefringence to be accounted for. The ASE photons are then received by the same polarization analyzers (PAs) as the quantum signals and is then detected by the single photon detectors. Active feedback from the single photon counts from the detectors are used to perform basis alignment by adjusting birefringence compensating waveplates in the PAs. 

\subsubsection{Q-LAN2 Experiment}

\begin{figure}
    \centering
    \includegraphics[width=.9\textwidth]{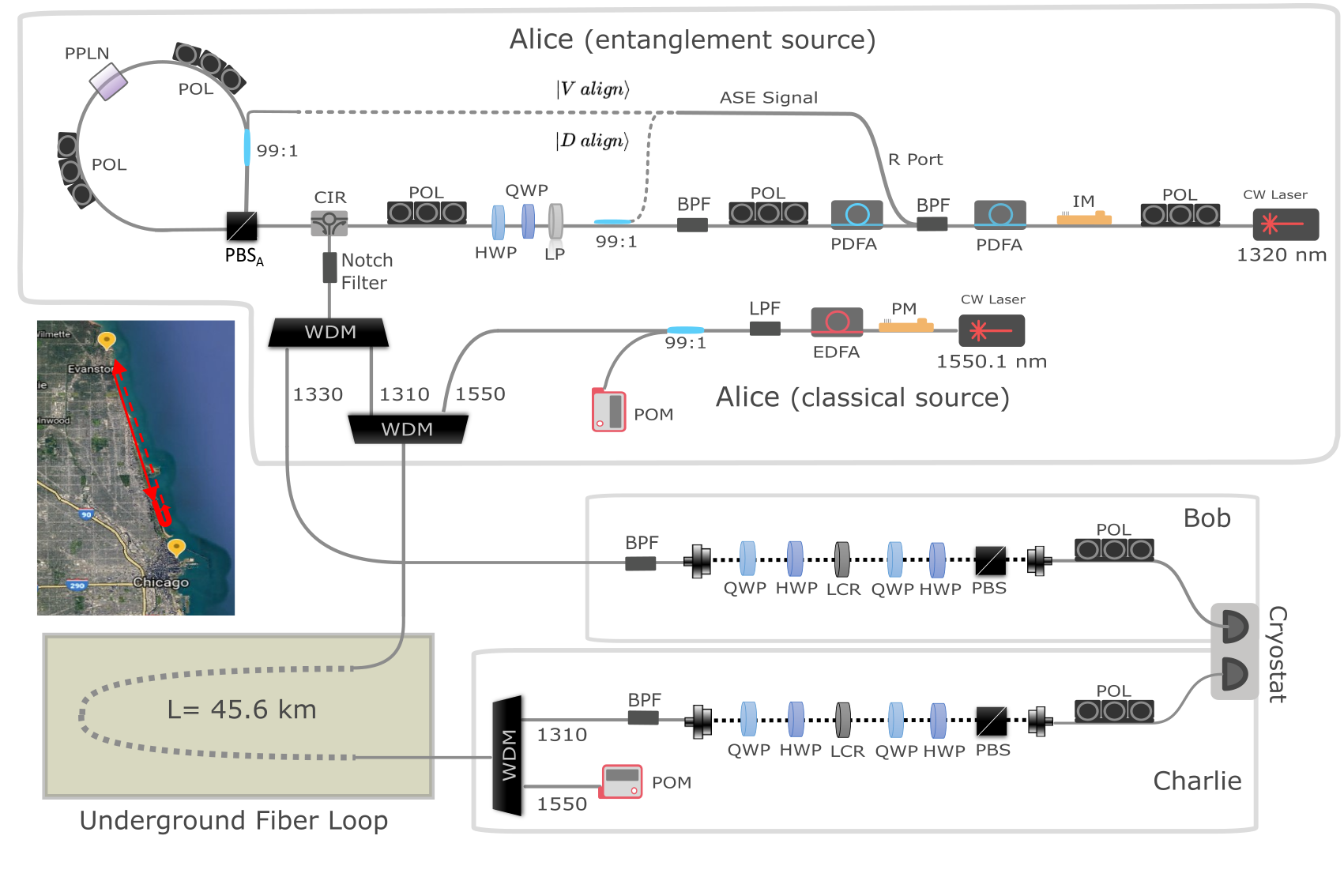}
    \caption{Q-LAN2 experimental design.}
    \label{fig:figure7}
\end{figure}

Figure~\ref{fig:figure7} shows a schematic of our experiment. We generate the polarization entangled quantum signals via cascaded second harmonic generation-spontaneous parametric down conversion (c-SHG-SPDC) in a single periodically poled lithium-niobate waveguide (PPLN)~\cite{arahira2011generation}. The cascaded second order interaction acts as a quasi-spontaneous four wave mixing process used in fiber-based sources of entanglement~\cite{li_optical-fiber_2005}, where the pump, signal, and idler wavelengths all occupy the same wavelength band. Since our cascaded second order nonlinear source is analogous to a fiber-based entanglement source, we easily adopt the built-in alignment signal design used in previous fiber-based sources of polarization entanglement~\cite{Wang2013, wang_multi-channel_2009} for this SPDC based source. The PPLN waveguide is phase matched for SHG of the 1320 nm pump pulse train at 417 MHz repetition rate with 80 ps pulse-width. We place the waveguide inside a polarization Sagnac loop to generate polarization entangled photon pairs. The pump light entering the loop is split by a polarizing beam splitter ($\text{PBS}_A$) into two counter-propagating directions. The c-SHG-SPDC generates broad-bandwidth quantum amplitudes for photon pairs centered around 1320 nm in both directions. Upon recombination at the PBS, the two photon state exits the Sagnac loop in a polarization entangled state $|\phi\rangle \propto |HH\rangle_A + e^{i \phi_{EPS}} |VV\rangle_A$, where $\phi_{EPS}$ is due to a relative phase between the H and V components of the pair generation pump. 

We then separate the signal/idler photons into bands using a standard coarse wavelength division multiplexer (CWDM). The CWDM outputs 20 nm wide bands with center wavelengths of 1310 and 1330 nm. The broadband spectrum of the c-SHG-SPDC allows the source to support distribution over the full O-band to easily scale the network to more users by carving the spectrum into more WDM output channels by either using the 1290 and 1350
CWDM bands or by further carving the 20 nm wide channels into multiple narrower channels.

For alignment, broadband ASE from a Praseodymium-doped fiber amplifier (PDFA) is multiplexed in to transmit polarized signals in the wavelength bands defined by the spectral filters in each fiber link connection. The alignment procedure consists of sending two sequential alignment signals, one with a vertical ($|V \ align \rangle$) polarization and the other with a signal for diagonal ($|diag \ align \rangle \propto \frac{1}{\sqrt{2}}(|H\rangle_A + e^{i \phi_{EPS}} |V\rangle_A)$ alignment, which the receiving nodes at Bob and Charlie use to orient their waveplates while counting single photons to align their polarization basis with that at Alice.  A 100 GHz DWDM filter centered at 1320 nm passes the pump wavelength (1320 nm) and rejects the broad ASE spectrum into another fiber. For H/V basis alignment the ASE signal is injected into one arm of the Sagnac loop in Alice's source to be reflected at PBS$_A$ such that it emerges vertically polarized (defined by Alice's PBS) and propagates through both fiber channels to the PAs. At $\text{PA}_{Bob}$ and $\text{PA}_{Charlie}$, we initially set the projective measurement waveplates to 0 degrees, with a liquid crystal retarder aligned at 0 degrees. The horizontal basis is then aligned via rotating the first QWP/HWP pair to minimize the single count rates in both Bob and Charlie's single photon detectors. 

The diagonal alignment signal is generated by injecting ASE via a 99:1 splitter into the source pump path such that the output signal carries the same relative weighting and phase of H and V polarizations as the entangled photons. At both PAs, the projection HWPs are set at 22.5 degrees to project onto the D basis while an automated search scans the voltages applied to the LCR, which adjusts the relative phase between H and V components until the singles count rates are minimized. After the optimal voltages are found for both PAs, the relative phase has been set to zero degrees resulting in the arrival of the symmetric Bell state $|\Phi \rangle_+ =\frac{1}{\sqrt{2}}(|H_B H_C\rangle+|V_B V_C\rangle)$ shared between Bob and Charlie, where $|H\rangle$ and $|V\rangle$ are defined at Alice's PBS. 

To demonstrate coexistence, we amplify C-band laser light at 1550.1 nm with an Erbium-doped fiber amplifier (EDFA) and multiplex it into the underground fiber to copropagate with the O-band quantum signal. We phase modulate the C-band light to broaden its spectrum which emulates a data channel and inhibits stimulated Brillouin scattering. At the receiver, we demultiplex out the C-band light while further filtering the remaining signal and idler photons with 100 GHz bandpass (BP) filters centered at 1306.5 and 1333.5 nm, respectively. We detect the photons with SNSPDs, which are followed by a time-tagging correlation detection system. We apply an electronic delay between the two channels to account for the fiber delay and perform coincidence measurements using a coincidence correlation time window of $\sim$0.5 ns, which is set such that the arriving photon pair pulse would not drift outside the window due to timing jitter from transmission over the fiber.

We send one photon of the polarization-entangled photon pair over a 45.6~km loop of underground installed fiber. The underground fiber link connects the Quantum Communications Laboratory at Northwestern University in Evanston to the Starlight Communications Facility located on the Northwestern Campus in Chicago (link distance of 22.8~km) where we loop the co-propagating light back to the Evanston laboratory for characterization with polarization analyzers and low-dark-count superconducting nanowire single photon detectors (SNSPDs).

\begin{figure*}
    \centering
    \includegraphics[width=\textwidth]{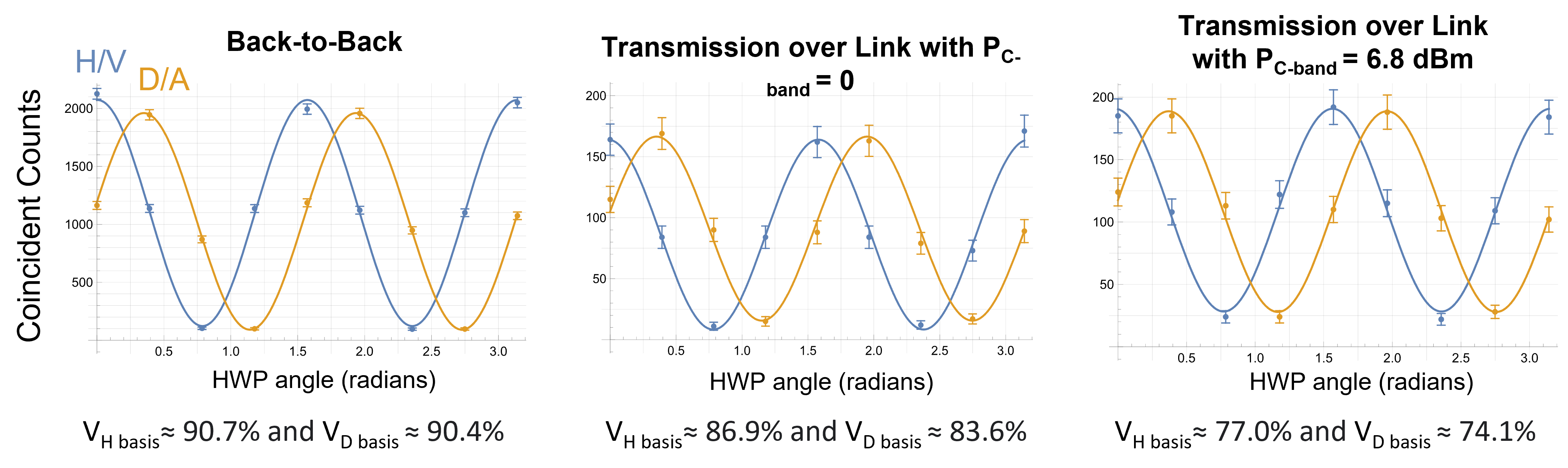}
    \caption{(a) Back-to-back entanglement visibility measurements. (b) Entanglement visibility after transmission over 45.6 km installed fiber link. (c) Entanglement visibility with 6.8 dBm copropagating C-band power.}
    \label{fig:figure8}
\end{figure*}

Polarization measurement apparatuses described above at Bob and Charlie first use the classical basis alignment signals from Alice to set their measurement basis to the same reference frame defined by Alice's source.  After basis alignment is done, two-photon interference (TPI) measurements are made to analyze the performance of entanglement distribution without (Figure~\ref{fig:figure8}(b)) and with (Figure~\ref{fig:figure8}(c)) copropagating classical light. 
TPI records coincidence-count fringes in two non-orthogonal basis states, and the resulting fringe visibility is a metric of entanglement quality. TPI fringes with V>70.7\% are consistent with violating Bell's inequalities and are strongly non-classical.  To record a TPI, the basis state measurement of one location is fixed (say in the H or D orientation) by appropriate alignment of the polarization control elements, while the basis state measurement of the other location is scanned along a Great Circle of the Poincaré sphere. 
For reference, Figure~\ref{fig:figure8}(a) shows the back-to-back TPI fringes, where visibilities are roughly 90\%. The low back-to-back visibilities are a result of increasing the source rates at the expense of visibility in order to achieve distribution over the lossy fiber. Figure~\ref{fig:figure8}(d) shows TPI fringes after the transmitted photon has propagated alongside 6.8 dBm of C-band launch power, where a visibility of 77\% is observed in the HV basis and 74\% in the DA basis. Both values are > 71\% and thus fall in the nonclassical regime of two-photon interference. 

Due to the lack of active temporal drift monitoring for electronically shifting the correlation window around the $\sim$100 ps pulse, we use a 500 ps correlation window to make sure the coincidence counts stay within the same time window.  Active monitoring and shifting of a narrower correlation window would significantly increase visibility. In the frequency domain, narrowing our frequency filtering even further to less than 5 GHz could presumably reduce Raman noise by another factor of $\sim$20, which would allow for even higher copropagating powers to be used. The powers used in our experiment clearly demonstrate that O-band quantum networks can coexist with the majority of modern classical communications channels in the C-band. Further improvement to the quantum source and the optimization of SNSPDs for the O-band (>80\% efficiency) would significantly increase the detection rates.

In summary, we have demonstrated that O-band/C-band quantum/classical wavelength allocation along with temporal and spectral filtering in coincidence detection are useful noise mitigation methods for coexistence scenarios in fiber-optic quantum networking. High visibility two photon interference fringes are demonstrated over 45.6 km of installed underground fiber with a copropagating $\sim$7 dBm classical launch power.  An O-band classical alignment signal is built into the entanglement source to align each node in a polarization entanglement network to the same polarization reference frame.  The combination of these two methods allows for robust polarization entanglement networks to be integrated into real world installed fiber networks.

\section{Conclusions and Future Work} \label{sec:conclusion}

The network architecture described in this paper, where SDN technology is used to decouple the control and data planes and perform network functions, could be easily be extended to include pair sources that are able to create entangled pairs over a broad wavelength rage. This in turn would allow for simultaneous distribution of entanglement from a single node to multiple users in a network by using wavelength-division multiplexing and by flexible assigning of the available frequency pairs to the different user pairs. Furthermore, to fully take advantage of this capability, solutions to achieving photon-packet interference from sources of different wavelength are needed (e.g., by using tuneable narrow band filters). This gives the network much more flexibility allowing a more efficient and dynamic quantum network.


The next steps on the way to the actual implementation of our quantum network will be distribution of the entangled photon between Fermilab (Q-LAN1) and Argonne National Lab (Q-LAN3) and demonstration of entanglement swapping operation at the Q-LAN1 for two independent entangled photon pair sources that are located at two distant Q-LAN1 labs within Fermilab. To perform this task it is crucial that the photons that are used for the Bell state measurement arrive at the same time. This requires high degree of synchronization between participating locations. To achieve this we will use the recently developed clock distribution system \cite{sync_paper} and characterized at the QLAN1. This system allows for synchronization at the few ps second level by sending optical clock signals using the same optical fiber as quantum signal, hence also demonstrating the ability of quantum-classical communication co-existence. We will also continue to pursue  co-existence demonstrators and quantum protocols at the Northwestern University locations (Q-LAN2) using different telecom wave bands for the classical and quantum signals. In the next step for the network deployment we will implement conversion between time-bin and polarization qubits and vice versa, which would allow us to perform teleportation and entanglement swapping between Q-LANs that use different types of entanglement. Further steps include connections to node-based hybrid systems such as superconducting resonators~\cite{qc-algo-sens-hep}, mechanical cavities or spins by direct transduction~\cite{lauk2020perspectives} or teleportation, growing to a quantum internet.


\section*{Acknowledgment}
IEQNET is funded by the Department of Energy’s Advanced Scientific Computing Research Transparent Optical Quantum Networks for Distributed Science program, but no government endorsement is implied.
The Caltech team also acknowledges partial funding support from the Department of Energy BES HEADS-QON Grant No. DE-SC0020376 on applications related to transduction, Department of Energy HEP-QuantiSED Grant No. DE-SC0019219, and the AQT Intelligent Quantum Networks and Technologies (INQNET) research program as well as  productive discussions with Matt Shaw and Boris Korzh of  the Jet Propulsion Laboratory, California Institute of Technology and Daniel Oblak, and Christoph Simon of the University of Calgary.   

\bibliographystyle{IEEEtran}
\bibliography{SPIEsubmission/SPIEbib.bib,SC21submission/bibliography.bib}

\newpage
\section*{Appendix}
\setcounter{figure}{0}
\renewcommand\thefigure{A.\arabic{figure}}
In this appendix we present sequence diagrams for the IEQNET control protocols described in the main text: (1) the quantum network resource and discovery protocol, and (2) the protocol for handling entanglement distribution requests.

\begin{figure*}
    \centering
    \includegraphics[width=.75\textwidth,trim={0 0 0 1cm},clip]{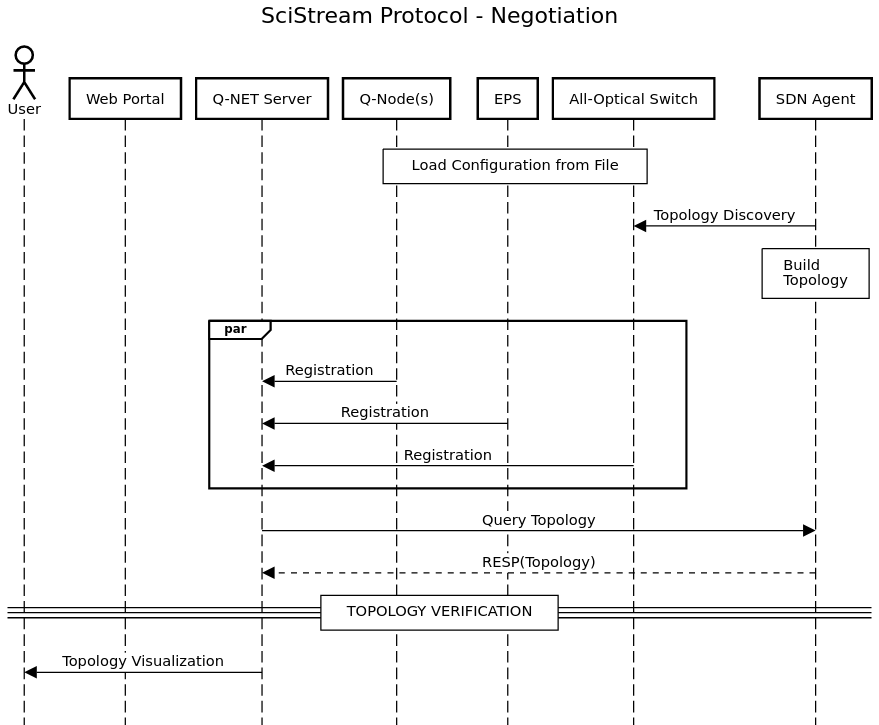}
    \caption{Quantum network resource and discovery protocol.}
    \label{fig:discovery}
\end{figure*}

As described in Section~\ref{sec:protocol}, the discovery protocol (see Figure~\ref{fig:discovery}) starts when each quantum networking resource (e.g., Q-Nodes, EPS, BSM-nodes, and switches) load their own configuration from file at initialization. The next step is for the SDN agent to discover the network topology through the SDN controller's southbound API (an interface used by SDN controllers to communicate with networking devices). Quantum network resources register to the Q-NET server by sending their features and connectivity information. The Q-NET server will request the topology from the SDN agent and will subsequently ask the SDN agent to verify the connectivity information provided by individual quantum resources. Once the topology has been verified, the Q-NET server will build a topology graph based on the updated topology and present it to users through the Web portal. 

\begin{figure*}
    \centering
    \includegraphics[width=.75\textwidth]{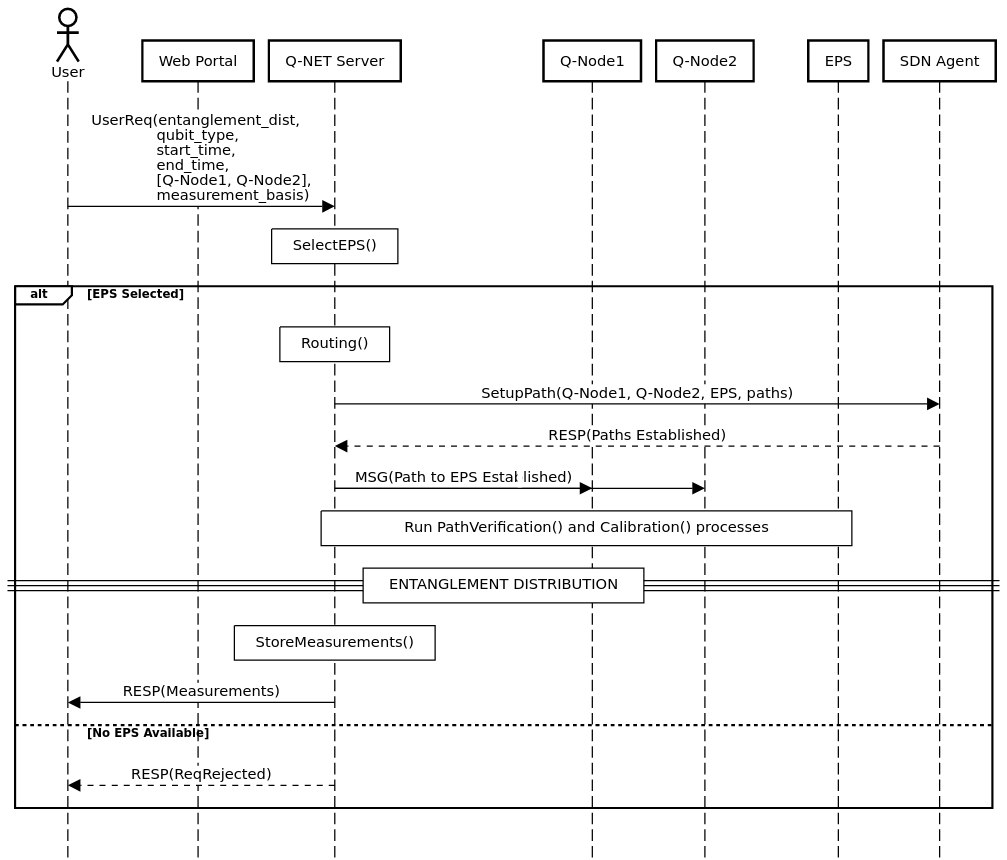}
    \caption{The protocol for handling entanglement distribution requests. The steps inside the \texttt{alt} box are only executed if the \texttt{SelectEPS()} procedure returns a successful ESP, otherwise the request is cancelled.}
    \label{fig:res-mgmt}
\end{figure*}

Figure~\ref{fig:res-mgmt} describes the the entanglement distribution protocol, which starts with a user requesting entanglement distribution between two Q-Nodes (e.g., Q-Node1 and Q-Node2). The Q-NET server will analyze this request and choose an EPS that meets the requirements specified by the user. Upon acceptance of the request, the Q-NET server will execute path routing and wavelength assignment and will establish the paths among involved entities via the SDN agent. Q-NET server will notify Q-Nodes 1 and 2 when paths are established and initiate path verification, which involves a series of active probes from EPS to Q-Nodes (and vice versa) using both classical and quantum light. After path verification, the Q-NET server will initiate calibration and optimization processes (as described in Section~\ref{sec:cp-design}) for the requested service. Once all entities send the READY signal to the Q-NET server, the entanglement distribution process starts. Q-Nodes will collect measurements until they have a long enough ebit string for the upper layer application. At that point they will send the END signal to the Q-NET server to stop entanglement distribution. Periodically during entanglement distribution, the Q-NET server will re-initiate the calibration and optimization processes. After the Q-NET server stops the EPS, all measurements will be stored at the Q-NET server and the user will be able to access them through the Web portal.

\end{document}